\documentclass[12pt]{article}
\usepackage[toc,page]{appendix}
\usepackage{setspace}
\usepackage{amssymb}
\usepackage{amsmath,amsfonts}
\usepackage{graphicx}
\usepackage{mathrsfs}
\usepackage[vcentermath]{youngtab}
\usepackage{color}
\numberwithin{equation}{section}
\usepackage{bbm}
\newcommand{\nn}{\nonumber\\}

\newcommand{\be}{\begin{equation}}
	\newcommand{\ee}{\end{equation}}
\newcommand{\bea}{\begin{eqnarray}}
	\newcommand{\eea}{\end{eqnarray}}

\newcommand{\eq}{&=&}

\newcommand{\ket}[1]{\left\lvert #1 \right\rangle}
\newcommand{\lb}{\left(}
\newcommand{\rb}{\right)}
\usepackage{multirow}
\newcommand{\half}{\frac{1}{2}}

\newcommand{\commute}[2]{\left[ #1 \, , \, #2 \right]}

\usepackage{array}
\usepackage{graphicx}
\usepackage{longtable}
\usepackage{color}
\usepackage{bbm}
\usepackage{appendix}

\newcommand{\bep}{\begin{picture}}
\newcommand{\eep}{\end{picture}}

\newcounter{YoungHeight}\newcounter{YoungWidth}

\newcounter{Mul1}\newcounter{Mul2}\newcounter{Mul3}\newcounter{Mul4}
\newcounter{A1}\newcounter{A2}
\newcounter{B3}
\newcounter{C3}\newcounter{C4}

\newcounter{T0}\newcounter{T1}

\newcounter{R0}

\newlength{\txtHShift}

\newlength{\txtWidth}

\newcommand{\Add}[3]{\setcounter{#1}{#2}\addtocounter{#1}{#3}}

\newcommand{\Length}[1]{#10}
\newcommand{\YoungScale}{}

\newcommand{\BlockA}[2]{{\YoungScale\bep(\Length{#1},\Length{#2}){\Add{A1}{#1}{1}\Add{A2}{#2}{1}}%
		\multiput(0,0)(10,0){\value{A1}}{\line(0,1){\Length{#2}}}\multiput(0,0)(0,10){\value{A2}}{\line(1,0){\Length{#1}}}%
		\setcounter{YoungHeight}{\Length{#2}}\setcounter{YoungWidth}{\Length{#1}}\eep}}

\newcommand{\YoungB}{\BlockA{2}{1}}

\newcommand{\YoungAA}{\BlockA{1}{2}}

\newcommand{\YoungBB}{\BlockA{2}{2}}

\newcommand{\YoungAAAA}{\BlockA{1}{4}}

\usepackage[normalem]{ulem}

\begin{document}

\begin{center}
{\large \bf Quasiconformal Group Approach to Higher Spin Algebras, their Deformations and Supersymmetric Extensions}

\vspace{1cm} 

{\bf Murat G\"unaydin\footnote{email: mgunaydin@psu.edu}}

\vspace{0.5cm} 

{\it 	Institute for Gravitation and the Cosmos \\
Physics Department,
	Pennsylvania State University \\
	University Park, PA 16802, USA}

\vspace{1cm} 
\end{center}
\begin{abstract} 
The quasiconformal method provides us with a unified approach to the construction of  minimal unitary representations (minrep)  of noncompact groups, their deformations as well as their supersymmetric extensions. We review the quasiconformal construction of the minrep of $SO(d,2)$,  its deformations and their applications to  unitary realizations of $AdS_{(d+1)}/CFT_d$ higher spin algebras and their deformations for arbitrary $d$ and supersymmetric extensions for $d\leq 6$. $AdS_{(d+1)}/CFT_d$  higher spin algebras, their deformations and supersymmetric extensions are given by the enveloping algebras of the quasiconformal realizations of the minrep, its deformations and supersymmetric extensions, respectively, and are in one-to-one correspondence with  massless conformal fields for arbitrary $d$  and massless conformal supermultiplets for $d\leq 6$. 
\end{abstract}

\vspace{2cm}
\begin{center}
{\it \small To appear in the "Proceedings of the International Workshop on Higher Spin Gauge Theories" , Institute of Advanced Studies, NTU, Singapore, \\ November 4-6, 2015}
\end{center}
\newpage

\section{Introduction}
The minimal unitary representation (minrep)  of a noncompact Lie group, as defined by Joseph  \cite{MR0342049}, is realized over an Hilbert space of functions that depend on the smallest number of variables possible. The mathematics literature on minreps of noncompact groups  is quite extensive.  We refer to  \cite{MR644845,MR1103588,binegar1991unitarization,MR1159103,MR1372999,MR1278630,MR1327538,MR1737731,MR2020550,MR2020551,MR2020552,Gover:2009vc,Kazhdan:2001nx} and the references therein. After the discovery of the novel  geometric quasiconformal realizations of non-compact groups\cite{Gunaydin:2000xr,Gunaydin:2005zz} 
a unified approach to their minreps was developed.  The minimal unitary realization is obtained by 
  quantization of  the geometric quasiconformal action. This was  first carried out  explicitly for the  largest exceptional group  $E_{8(8)}$ \cite{Gunaydin:2001bt} which is the U-duality group of maximal supergravity in three dimensions. The minrep  of  the $3d$  U-duality group $E_{8(-24)}$ of the exceptional supergravity \cite{Gunaydin:1983rk} via the quasiconformal approach was constructed in \cite{Gunaydin:2004md}.
A unified approach to the construction of the minreps of noncompact groups by quantizing their quasiconformal actions was formulated in \cite{Gunaydin:2006vz}. The results of \cite{Gunaydin:2006vz}  extend to the minreps of noncompact superalgebras $\mathfrak{g}$ whose even subalgebras are of the form $\mathfrak{h}\oplus \mathfrak{sl}(2,\mathbb{R})$ where $\mathfrak{h}$ is compact. The quasiconformal construction of the minreps of non-compact Lie algebras  $SU(n,m|p+q)$ and $OSp(2N^*|2M)$ were subsequently developed in \cite{Fernando:2009fq,Fernando:2010dp,Fernando:2010ia}\footnote{We will use the  notation for labelling Lie (super)groups and Lie (super)algebras interchangeably throughout the paper.} 

The quasiconformal construction of the minrep of   $AdS_5/CFT_4$  group  $SU(2,2)$ and  its supersymmetric extensions $SU(2,2|N)$ were studied in  \cite{Fernando:2009fq}. 
The minrep of $SU(2,2)$  describes a massless scalar field in $4d$. It admits a one parameter ($\zeta$) family  of deformations labelled by helicity, which can be continuous\footnote{The parameter $\zeta$ is twice the helicity.}. For a positive (negative)  integer value of  $\zeta$,  the deformed minrep describes a massless conformal field transforming in $\left(0 \,,\, \frac{\zeta}{2} \right)$ $\lb \lb -\frac{\zeta}{2},0 \rb\rb$ representation of the Lorentz group. The minrep and its  deformations  for integer $\zeta$ are isomorphic to the doubleton representations  of $SU(2,2)$ that were constructed using covariant twistorial oscillators\cite{Gunaydin:1984fk,Gunaydin:1998sw,Gunaydin:1998jc}. Similarly, the minrep  of $SU(2,2\,|\,N)$ and its deformations for integer $\zeta$  are isomorphic to  the  doubleton supermultiplets that were studied in \cite{Gunaydin:1984fk,Gunaydin:1998sw,Gunaydin:1998jc}.The minimal unitary supermultiplet of
$PSU(2,2|4)$  is  the $N=4$  Yang-Mills supermultiplet in $d=4$ that was first constructed as a CPT self-conjugate .
doubleton supermuliplet in \cite{Gunaydin:1984fk}.


The minrep of $AdS_7/CFT_6$ group  $SO(6,2)=SO^*(8)$ and its deformations were constructed  in \cite{Fernando:2010dp}. The deformations of the minrep of $SO(6,2)$ are labelled by the "spin" $t$ of an $SU(2)$ subgroup . The $SU(2)$ spin $t$  is the analog of helicity in six dimensions. The minrep of $SO^*(8)$ and its deformations  describe massless  conformal fields in six dimensions.
These results were then extended to $6d$ superconformal algebras  $OSp(8^*|2N)$  \cite{Fernando:2010dp,Fernando:2010ia}. The minimal unitary supermultiplet  of $OSp(8^*|4)$ describes  the massless conformal $(2,0)$ supermultiplet in $6d$, which was first constructed as a doubleton supermultiplet in \cite{Gunaydin:1984wc}.

The quasiconformal realizations of minreps of noncompact groups are in general nonlinear involving operators that are cubic and quartic in the coordinates and momenta. However for symplectic groups quasiconformal construction of the minrep coincides with  the oscillator construction involving bilinears of  twistorial oscillators\cite{Gunaydin:2006vz}. Hence the minrep of $AdS_4/CFT_3$ group $Sp(4,\mathbb{R})$ is simply the scalar singleton of Dirac and it admits a single deformation which is the spinor singleton. They were referred to as the remarkable representations of $AdS_4$ group  by Dirac \cite{dirac1963remarkable}.  The singleton supermultiplets of  $AdS_4$ superalgebras $OSp(2n|4,\mathbb{R})$ , in particular the  $N=8$ superalgebra $OSp(8|4,\mathbb{R})$,  were first constructed  in \cite{Gunaydin:1983cc,Gunaydin:1983yj} using the oscillator methods developed earlier \cite{Gunaydin:1981yq,Bars:1982ep}. Oscillator construction of general supermultiplets of $OSp(N|4,\mathbb{R})$  was later given in \cite{Gunaydin:1984wc,Gunaydin:1988kz}.

All the massless higher spin representations of $AdS_4$ group $SO(3,2)$ occur in the tensor product  of two singleton representations \cite{Flato:1978qz}. The  higher spin theories in $AdS_4 $ were studied extensively by Fronsdal and collaborators \cite{flato1981quantum,Fronsdal:1981gq,Flato:1980we,Angelopoulos:1980wg}. In the 1980s Fradkin and Vasiliev pioneered the study of higher spin theories involving fields containing infinitely  many spins $0 \leq s < \infty $ \cite{Fradkin:1986qy,Konshtein:1988yg}. Much work has been done on higher spin theories in the intervening years, in particular after the work of \cite{Klebanov:2002ja} who conjectured that  Vasiliev's higher spin  theory\cite{Vasiliev:1992av} in $AdS_4$ is dual to $O(N)$ vector model in $3d$. The conjecture of \cite{Klebanov:2002ja} was  checked by Giombi and Yi   by  calculating some of the three point functions of higher spin currents in the bulk  and matching them with those of free and critical $O(N)$ vector models in $3d$ \cite{Giombi:2009wh,Giombi:2010vg}.
For  reviews on higher spin theories  we refer to \cite{Vasiliev:1995dn,Vasiliev:1999ba,Bekaert:2005vh,Iazeolla:2008bp,Sagnotti:2011qp,Giombi:2012ms,Didenko:2014dwa} and references therein. 

That the Fradkin-Vasiliev higher spin algebra in $AdS_4$ \cite{Fradkin:1986qy} is   the enveloping algebra of the singletonic realization of $Sp(4,\mathbb{R})$  was first pointed out in \cite{Gunaydin:1989um}. Again in \cite{Gunaydin:1989um} it was suggested that the higher spin algebras of Fradkin-Vasiliev type  in $AdS_5$ and $AdS_7$  can similarly be obtained from the doubletonic realizations of $SU(2,2)$ and $SO(6,2)$ , respectively.  
Higher spin superalgebras in $AdS_4$, $AdS_5$ and $AdS_7$ could then be realized  as enveloping algebras of the singletonic  realization of $OSp(N/4,\mathbb{R})$ and doubletonic realizations of  $SU(2,2|N)$  and $OSp(8^*|2N)$\cite{Gunaydin:1989um}. Conformal higher spin superalgebras in four dimensions were studied  in \cite{Fradkin:1989yd}.
Higher spin theories and supersymmetric extensions in $AdS_5$ and $AdS_7$ were studied later  in \cite{Sezgin:2001zs,Sezgin:2001ij,Sezgin:2001yf,Sezgin:2002rt} using the doubletonic  realizations of  \cite{Gunaydin:1984wc,Gunaydin:1998jc,Gunaydin:1998sw,Fernando:2001ak}.  Higher spin superalgebras in higher dimensions  were also studied in \cite{Vasiliev:2004cm}. 

Vasiliev pointed out that to obtain the standart bosonic higher spin algebra in $AdS_{(d+1)}$, one has to quotient the enveloping algebra of $SO(d,2)$  by the ideal that annihilates  the scalar ``singleton'' representation \cite{Vasiliev:1999ba}. Later Eastwood identified this ideal to be the Joseph ideal \cite{Eastwood:2002su}.
This is consistent with the observation that  $AdS_4/CFT_3$ higher spin algebra is given by the enveloping algebra of the singletonic realization of $SO(3,2)$\cite{Gunaydin:1989um}  since  the Joseph ideal vanishes identically for the singleton. However  the Joseph ideal does not vanish identically as operators for the doubletonic realizations of $SO(4,2)$ and  $SO(6,2)$ in terms of covariant twistorial oscillators. This is where the importance of quasiconformal approach to the construction of higher spin algebras and superalgebras and their deformations becomes manifest.  The Joseph ideal vanishes identically as operators in the quasiconformal construction of the minrep of $SO(d,2)$ and hence its  enveloping algebra leads directly to the higher spin algebra without the need for quotienting. This approach also allows one to define deformations and supersymmetric extensions of higher spin algebras. 

The vanishing of the Joseph ideal  for the quasiconformal realization of minreps of $SO(4,2)$ and  $SO(6,2)$ was shown in \cite{Govil:2013uta,Govil:2014uwa}. Hence  their enveloping algebras lead directly to unitary realizations of the bosonic $AdS_5/CFT_4$ and $AdS_7/CFT_6$  higher spin algebras, respectively. The enveloping algebras of the deformed minreps of $SU(2,2)$ and of $SO^*(8)$ and their supersymmetric extensions yield  infinite families of $AdS_5/CFT_4$ and $AdS_7/CFT_6$ higher spin algebras and superalgebras, respectively \cite{Govil:2013uta,Govil:2014uwa}.

These results were extended to $AdS_6/CFT_5$ higher spin algebra in \cite{Fernando:2014pya}.  The minrep of $SO(5,2)$ admits a  unique deformation which describes a conformally massless spinor field in $5d$. In five dimensions there exists a unique simple conformal superalgebra , namely $F(4)$ with the even subalgebra $SO(5,2)\oplus SU(2)$. The minimal unitary supermultiplet of $F(4)$ decomposes into the deformed minrep and two copies of the minrep. Its enveloping algebra defines the unique higher spin superalgebra in $AdS_6$.

Extension of the above results to higher dimensions was given in \cite{Fernando:2015tiu} where it was shown that   the minrep of $SO(d,2)$ and its deformations are in one-to-one correspondence with massless conformal fields in $d$ dimensional spacetimes. For the quasiconformal realization of the minrep of $SO(d,2)$  generators of the Joseph ideal vanish  and its enveloping algebra  yields directly the  $AdS_{(d+1)}/CFT_d$ higher spin algebra. The enveloping algebra of a deformation of the minrep leads to a deformed higher spin algebra.   In odd dimensions there exists a unique deformation.  In even dimensions there exist infinitely many deformations whose enveloping algebras define an infinite family of higher spin algebras.

Below we will review the quasiconformal realization of $SO(d,2)$ and its quantization that leads directly to the minimal unitary representation of $SO(d,2)$. We shall then discuss the deformations of the minrep and show that there is a one-to-one correspondence between the minrep and its deformations and massless conformal fields in $d$ dimensional Minkowskian spacetimes. This will be followed by a review of the application of these results to higher spin algebras, their deformations and supersymmetric extensions.

\section{Geometric realization of $SO(d,2)$ as a quasiconformal group}
\label{sec:geomSO(d,2)}

The geometric quasiconformal realization of the anti-de Sitter (conformal) group $SO(d,2)$ in $(d+1)$ $(d)$ dimensions  that was given  in \cite{Gunaydin:2005zz} is based on the  5-grading of the Lie algebra $\mathfrak{so}(d,2)$ with respect to its maximal rank subalgebra $\mathfrak{so}(1,1) \oplus \mathfrak{so}(d-2) \oplus \mathfrak{so}(2,1)$ :

\be	\mathfrak{so}(d,2)
	= \mathbf{1}^{(-2)} \oplus
	\left( \mathbf{d-2} , \mathbf{2} \right)^{(-1)} \oplus
	\left[ \,
	\Delta \oplus
	\mathfrak{so}(d-2) \oplus
	\mathfrak{su}(1,1)
	\, \right]^0 \oplus
	\left( \mathbf{d-2} , \mathbf{2} \right)^{(+1)} \oplus
	\mathbf{1}^{(+2)} \nonumber \ee
where the generator that determines the five grading is denoted as $\Delta$. We should stress that $SO(d-2)$ is the simple part of the little group of massless particles in $d$-dimensional Minkowski space-time. The generators of  quasiconformal group action of $SO(d,2)$ are realized  as nonlinear differential operators acting on a $\left( 2 d -3 \right)$-dimensional space $\mathcal{T}$ with coordinates  $\mathcal{X}= \left( X^{i,a} , x \right)$, where $X^{i,a}$ transform in the $(d-2,2)$ representation of $\mathfrak{so}(d-2) \oplus \mathfrak{su}(1,1)$ subalgebra, with $i=1,\dots,d-2$ and $a=1,2$, and $x$ is a singlet coordinate.

Labelling  the generators with grade $-2,-1,0,+1$ and $+2$ as $K_- , U_{i,a} 
, \, [ \Delta \oplus \mathcal{L}_{ij} \oplus \mathcal{M}_{ab} ], \,
\widetilde{U}_{i,a}$ and $
K_{+}$, respectively,  their quasiconformal realizations have the form:
\begin{equation} \label{QCGaction}
	\begin{split}
		K_+
		&= \frac{1}{2} \left( 2 x^2 - \mathcal{I}_4 \right) \frac{\partial}{\partial x}
		- \frac{1}{4} \frac{\partial \mathcal{I}_4}{\partial X^{i,a}}
		\eta^{ij} \epsilon^{ab} \frac{\partial}{\partial X^{j,b}}
		+ x \, X^{i,a} \frac{\partial}{\partial X^{i,a}}
		\\
		U_{i,a}
		&= \frac{\partial}{\partial X^{i,a}}
		- \eta_{ij} \epsilon_{ab} \, X^{j,b} \frac{\partial}{\partial x}
		\\
		\mathcal{L}_{ij}
		&= \eta_{ik} X^{k,a} \frac{\partial}{\partial X^{j,a}}
		- \eta_{jk} X^{k,a} \frac{\partial}{\partial X^{i,a}}
		\\
		\mathcal{M}_{ab}
		&= \epsilon_{ac} X^{i,c} \frac{\partial}{\partial X^{i,b}}
		+ \epsilon_{bc} X^{i,c} \frac{\partial}{\partial X^{i,a}}
		\\
		K_-
		&= \frac{\partial}{\partial x}
		\qquad , \qquad
		\Delta
		= 2 \, x \frac{\partial}{\partial x}
		+ X^{i,a} \frac{\partial}{\partial X^{i,a}}
		\qquad , \qquad
		\widetilde{U}_{i,a}
		= \commute{U_{i,a}}{K_+}
	\end{split}
\end{equation}
where $\mathcal{L}_{ij}$ and $\mathcal{M}_{ab}$ are the generators of $SO(d-2)$ and $SU(1,1)$ subgroups, respectively,($i,j,k,l=1,\dots,d-2$ ; $a,b,c,d=1,2$) and  $\epsilon^{ab}$ is the inverse symplectic tensor, such that $\epsilon^{ab} \epsilon_{bc} = {\delta^a}_c$ . $ \mathcal{I}_4 (X)$
denotes the quartic polynomial of the coordinates $X^{i,a}$
\begin{equation}
	\mathcal{I}_4 (X)
	= \eta_{ij} \eta_{kl} \epsilon_{ac} \epsilon_{bd}
	X^{i,a} X^{j,b} X^{k,c} X^{l,d}
\end{equation}
which is an invariant of $SO(d-2) \times SU(1,1)$ subgroup\footnote{Note that we have the isomorphisms $SU(1,1) \equiv SO(2,1) \equiv Sp(2,\mathbb{R})$.}. In the above expression, $\epsilon_{ab}$ is the symplectic invariant tensor of $SU(1,1)$ and $\eta_{ij}$ is the invariant metric of $SO(d-2)$ in the fundamental representation, which we choose as $\eta_{ij} = -\delta_{ij}$ to be consistent with the  conventions of \cite{Gunaydin:2005zz}.

The grade +1 generators $\widetilde{U}_{i,a}$ are obtained by substituting the expression for the quartic invariant in the above equations: 
\begin{equation}
	\begin{split}
		\widetilde{U}_{i,a} &= \commute{U_{i,a}}{K_+} \\
		&= \eta_{ij} \epsilon_{ad}
		\left(
		\eta_{kl} \epsilon_{bc} X^{j,b} X^{k,c} X^{l,d}
		- x \, X^{j,d}
		\right)
		\frac{\partial}{\partial x}
		+  x \frac{\partial}{\partial X^{i,a}}
		\\
		& \quad
		- \eta_{ij} \epsilon_{ab} \, X^{j,b} X^{l,c}
		\frac{\partial}{\partial X^{l,c}}
		- \epsilon_{ad} \eta_{kl} \, X^{l,d} X^{k,c}
		\frac{\partial}{\partial X^{i,c}}
		\\
		& \quad
		+ \epsilon_{ad} \eta_{ij} \, X^{l,d} X^{j,b}
		\frac{\partial}{\partial X^{l,b}}
		+ \eta_{ij} \epsilon_{bc} X^{j,b} X^{l,c}
		\frac{\partial}{\partial X^{l,a}}
	\end{split}
\end{equation}
The generators of $SO(d,2)$  satisfy the following  commutation relations:
\begin{subequations}
	\label{eq:sod2algebra}
	\begin{equation}
		\begin{split}
			\commute{\mathcal{L}_{ij}}{\mathcal{L}_{kl}}
			&= \eta_{jk} \mathcal{L}_{il} - \eta_{ik} \mathcal{L}_{jl} - \eta_{jl} \mathcal{L}_{ik} + \eta_{il} \mathcal{L}_{jk}
			\\
			\commute{\mathcal{M}_{ab}}{\mathcal{M}_{cd}}
			&= \epsilon_{cb} \mathcal{M}_{ad} + \epsilon_{ca} \mathcal{M}_{bd}
			+ \epsilon_{db} \mathcal{M}_{ac} + \epsilon_{da} \mathcal{M}_{bc}
		\end{split}
	\end{equation}
	\begin{equation}
		\begin{split}
			\commute{\Delta}{K_\pm}
			&= \pm 2 \, K_\pm
			\qquad \qquad \qquad
			\commute{K_-}{K_+}
			= \Delta
			\\
			\commute{\Delta}{U_{i,a}}
			&= - U_{i,a}
			\qquad \qquad \qquad \quad
			\commute{\Delta}{\widetilde{U}_{i,a}}
			= \widetilde{U}_{i,a}
			\\
			\commute{U_{i,a}}{K_+}
			&= \widetilde{U}_{i,a}
			\qquad \qquad \qquad \quad
			\commute{\widetilde{U}_{i,a}}{K_-}
			= - U_{i,a}
			\\
			\commute{U_{i,a}}{U_{j,b}}
			&= 2 \, \eta_{ij} \epsilon_{ab} \, K_-
			\qquad \qquad
			\commute{\widetilde{U}_{i,a}}{\widetilde{U}_{j,b}}
			= 2 \, \eta_{ij} \epsilon_{ab} \, K_+
		\end{split}
	\end{equation}
	\begin{equation}
		\begin{split}
			\commute{\mathcal{L}_{ij}}{U_{k,a}}
			&= \eta_{jk} U_{i,a} - \eta_{ik} U_{j,a}
			\qquad \qquad
			\commute{\mathcal{L}_{ij}}{\widetilde{U}_{k,a}}
			= \eta_{jk} \widetilde{U}_{i,a} - \eta_{ik} \widetilde{U}_{j,a}
			\\
			\commute{\mathcal{M}_{ab}}{U_{i,c}}
			&= \epsilon_{cb} U_{i,a} + \epsilon_{ca} U_{i,b}
			\qquad \qquad
			\commute{\mathcal{M}_{ab}}{\widetilde{U}_{i,c}}
			= \epsilon_{cb} \widetilde{U}_{i,a} + \epsilon_{ca} \widetilde{U}_{i,b}
		\end{split}
	\end{equation}
	\begin{equation}
		\commute{U_{i,a}}{\widetilde{U}_{j,b}}
		= \eta_{ij} \epsilon_{ab} \, \Delta
		- 2 \, \epsilon_{ab} \mathcal{L}_{ij}
		- \eta_{ij} \mathcal{M}_{ab}
	\end{equation}
\end{subequations}

The above nonlinear realization has a geometric interpretation as the invariance group of a light-cone with respect to a quartic distance function in the space $\mathcal{T}$. The quartic   norm of a vector $\mathcal{X}= (X,x)$ in the $(2d-3)$-dimensional space $\mathcal{T}$ is given as 
\begin{equation}
\mathcal{N}_4 \left( \mathcal{X} \right)
= \mathcal{I}_4 \left( X \right) + 2 \, x^2 \,.
\end{equation}
and the quartic distance function between any two points $\mathcal{X}$ and $\mathcal{Y}$ in $\mathcal{T}$ is then defined as follows: \cite{Gunaydin:2001bt,Gunaydin:2006vz}
\begin{equation}
	d \left( \mathcal{X} , \mathcal{Y} \right)
	= \mathcal{N}_4  \left( \delta \left( \mathcal{X} , \mathcal{Y} \right) \right)
\end{equation}
where the ``symplectic'' difference $\delta \left( \mathcal{X} , \mathcal{Y} \right)$ is defined as 
\begin{equation}
	\delta \left( \mathcal{X} , \mathcal{Y} \right)
	= \left( X^{i,a} - Y^{i,a} \,,\, x - y - \eta_{ij} \epsilon_{ab} \, X^{i,a} Y^{j,b} \right)=-  	\delta \left( \mathcal{Y} , \mathcal{X} \right)\,.
\end{equation}
The lightlike separations between any two points with respect to this quartic distance function are left invariant under the quasiconformal group action of $SO(d,2)$  on $\mathcal{T}$.

\section{Minimal unitary representation of $SO(d,2)$ from its quasiconformal realization}
\label{sec:minrepSO(d,2)}

The "quantization" of the  geometric quasiconformal realization of a noncompact group leads directly to its minimal unitary representation \cite{Gunaydin:2001bt,Gunaydin:2004md,Gunaydin:2005zz,Gunaydin:2006vz,Fernando:2015tiu}. To "quantize" the quasiconformal realization of  $SO(d,2)$, we split the $2(d-2)$ variables $X^{i,a}$ defined in section \ref{sec:geomSO(d,2)} into $(d-2)$ coordinates $X^i$ and $(d-2)$ conjugate momenta $P_i$ as
\begin{equation}
X^i = X^{i,1}
\qquad \qquad \qquad
P_i = \eta_{ij} \, X^{j,2}
\end{equation}
and introduce a  momentum $p$ conjugate to the singlet coordinate $x$ as well. Furthermore we treat them as quantum mechanical operators satisfying the canonical commutation relations
\begin{equation}
\commute{X^i}{P_j} = i \delta^i_j
\qquad \qquad \qquad
\commute{x}{p} = i \,.
\end{equation}

Instead of the coordinates  $X^i$ and momenta $P_i$ it will be convenient to work with  bosonic oscillator annihilation operators $a_i$ and creation operators $a_i^\dag$, defined as
\begin{equation}
a_i = \frac{1}{\sqrt{2}} \left( X^i + i \, P_i \right)
\qquad \qquad \qquad
a_i^\dag = \frac{1}{\sqrt{2}} \left( X^i - i \, P_i \right)
\end{equation}
They satisfy the commutation relations
\begin{equation}
\commute{a_i}{a_j^\dag} = \delta_{ij}
\qquad \qquad \qquad
\commute{a_i}{a_j} = \commute{a_i^\dag}{a_j^\dag} = 0 \,.
\end{equation}

For the minimal unitary realization of $\mathfrak{so}(d,2)$ the  5-grading
\begin{equation}
\mathfrak{so}(d,2)
= \mathfrak{g}^{(-2)} \oplus \mathfrak{g}^{(-1)} \oplus
\left[ \,
\Delta \oplus \mathfrak{so}(d-2)_L \oplus \mathfrak{su}(1,1)_M
\, \right] \oplus
\mathfrak{g}^{(+1)} \oplus \mathfrak{g}^{+2)} \nonumber
\end{equation}
is determined by the $SO(1,1)$ generator
\begin{equation}
\Delta = \frac{1}{2} \left( x p + p x \right) \,.
\label{delta}
\end{equation}

The generators of  $\mathfrak{su}(1,1)_M \subset \mathfrak{g}^{(0)}$ are  bilinears of the  bosonic oscillators:
\begin{equation}
M_+ = \frac{1}{2} a_i^\dag a_i^\dag
\qquad
M_- = \frac{1}{2} a_i a_i
\qquad
M_0 = \frac{1}{4} \left( a_i^\dag a_i + a_i a_i^\dag \right)
\label{SU(1,1)M}
\end{equation}
and satisfy
\begin{equation}
\commute{M_-}{M_+} = 2 \, M_0
\qquad \qquad
\commute{M_0}{M_\pm} = \pm \, M_\pm \,.
\end{equation}
Its quadratic Casimir  $\mathcal{M}^2$ is given by :
\begin{equation}
\mathcal{C}_2 \left[ \mathfrak{su}(1,1)_M \right]
= \mathcal{M}^2
= {M_0}^2 - \frac{1}{2} \left( M_+ M_- + M_- M_+ \right)
\end{equation}
The subalgebra $\mathfrak{so}(d-2)_L \subset \mathfrak{g}^{(0)}$  is also realized as bilinears of the  bosonic oscillators:
\begin{equation}
L_{ij} = i \left( a_i^\dag a_j - a_j^\dag a_i \right)
\end{equation}
and satisfy the commutation relations
\begin{equation}
\commute{L_{ij}}{L_{kl}}
= i \left( \delta_{jk} L_{il} - \delta_{ik} L_{jl} - \delta_{jl} L_{ik} + \delta_{il} L_{jk} \right) \,.
\end{equation}
The Casimir $\mathcal{L}^2$ of $\mathfrak{so}(d-2)_L$ 
\begin{equation}
\mathcal{C}_2 \left[ \mathfrak{so}(d-2)_L \right]
= \mathcal{L}^2
= L_{ij} L_{ij}
\end{equation}
is related to the Casimir of $\mathfrak{su}(1,1)_M$ as follows:
\begin{equation}
\mathcal{L}^2 = 8 \, \mathcal{M}^2 - \frac{1}{2} \left( d - 2 \right) \left( d - 6 \right) \,.
\end{equation}
The  generator in $\mathfrak{g}^{(-2)}$ is defined as
\begin{equation}
K_- = \frac{1}{2} x^2 \,.
\label{K-}
\end{equation}
The generators $(U_i , U_i^\dagger)$ in grade $-1$ subspace are realized as bilinears of $x$ and the  bosonic oscillators 
\begin{equation}
U_i = x \, a_i
\qquad \qquad \qquad
U_i^\dag = x \, a_i^\dag \,.
\label{Grade-1Bosonic}
\end{equation}
They close into $K_-$ under commutation and  form a Heisenberg subalgebra
\begin{equation}
\commute{U_i}{U_j^\dag} = 2 \, \delta_{ij} \, K_-
\qquad \qquad
\commute{U_i}{U_j} = \commute{U_i^\dag}{U_j^\dag} = 0
\end{equation}
with  $K_-= \frac{1}{2}x^2$ playing the role of  "central charge".

The quartic invariant $\mathcal{I}_4$ of $SO(d-2)_L \times SU(1,1)_M$ subgroup that enters the quasiconformal realization goes over to  a linear function of the quadratic Casimir of $SO(d-2)_L \times SU(1,1)_M$  after "quantization". In particular the grade $+2$ generator $K_+$ becomes:
\begin{equation}
K_+
= \frac{1}{2} p^2
+ \frac{1}{x^2} \mathcal{G}
\end{equation}
with $\mathcal{G}$ given by the Casimir $\mathcal{L}^2$:
\begin{equation}
\mathcal{G}
= \frac{1}{4} \mathcal{L}^2 + \frac{1}{8} \left( d - 3 \right) \left( d - 5 \right)
= 2 \mathcal{M}^2 + \frac{3}{8}
\label{IsotonicCouplingConstant}
\end{equation}
The  generators $W_i , W_i^\dagger $  in grade $+1$ subspace are given by  the commutators of the  grade $-1$ generators with  $K_+$:
\begin{equation}
\begin{split}
W_i
= - i \commute{U_i}{K_+} 
&= p \, a_i - \frac{i}{x} \left[ \frac{1}{2} \left( d - 3 \right) \, a_i + i \, L_{ij} \, a_j \right]
\\
W_i^\dag = - i \commute{U_i^\dag}{K_+}
&= p \, a_i^\dag - \frac{i}{x} \left[ \frac{1}{2} \left( d - 3 \right) \, a_i^\dag + i \, L_{ij} \, a_j^\dag \right] \,.
\end{split}
\end{equation}
The positive grade  generators form a Heisenberg algebra as well:
\begin{equation}
\commute{W_i}{W_j^\dag} = 2 \, \delta_{ij} \, K_+
\qquad \qquad
\commute{W_i}{W_j} = \commute{W_i^\dag}{W_j^\dag} = 0
\end{equation}
with the generator $K_+$ playing the role of  central charge. The commutators of grade $-2$ and grade $+1$ generators close into grade $-1$ subspace:
\begin{equation}
\commute{W_i}{K_-} = - i \, U_i
\qquad \qquad
\commute{W_i^\dag}{K_-} = - i \, U_i^\dag
\end{equation}


The generators $K_{\pm}$ and $\Delta$ form a distinguished $\mathfrak{su}(1,1)$ subalgebra labelled  as $\mathfrak{su}(1,1)_K$:
\begin{equation}
\commute{K_-}{K_+} = i \, \Delta
\qquad \qquad
\commute{\Delta}{K_\pm} = \pm 2 i \, K_\pm
\end{equation}
Its quadratic Casimir 
\begin{equation}
\mathcal{C}_2 \left[ \mathfrak{su}(1,1)_K \right]
= \mathcal{K}^2
= \Delta^2 - 2 \left( K_+ K_- + K_- K_+ \right)
\end{equation}
is related to the quadratic Casimir of $\mathfrak{so}(d-2)_L$ (and that of $\mathfrak{su}(1,1)_M$) as follows
\begin{equation}
\mathcal{K}^2 = - \frac{1}{2} \mathcal{L}^2 - \frac{1}{4} \left( d - 2 \right) \left( d - 6 \right) = - 4 \, \mathcal{M}^2 \,.
\end{equation}

Grade $\pm 1$  generators transform in the $(d-2,2)$ representation  $SO(d-2)_L \times SU(1,1)_M$:
\begin{equation}
\begin{aligned}
\commute{M_0}{U_i} &= - \frac{1}{2} \, U_i
\\
\commute{M_+}{U_i} &= - U_i^\dag
\\
\commute{M_-}{U_i} &= 0
\\
\commute{L_{ij}}{U_k} &= i \left( \delta_{jk} U_i - \delta_{ik} U_j \right)
\end{aligned}
\qquad \qquad \qquad
\begin{aligned}
\commute{M_0}{W_i} &= - \frac{1}{2} \, W_i
\\
\commute{M_+}{W_i} &= - W_i^\dag
\\
\commute{M_-}{W_i} &= 0
\\
\commute{L_{ij}}{W_k} &= i \left( \delta_{jk} W_i - \delta_{ik} W_j \right)
\end{aligned}
\end{equation}
\begin{equation}
\begin{aligned}
\commute{M_0}{U_i^\dag} &= \frac{1}{2} \, U_i^\dag
\\
\commute{M_+}{U_i^\dag} &= 0
\\
\commute{M_-}{U_i^\dag} &= U_i
\\
\commute{L_{ij}}{U_k^\dag} &= i \left( \delta_{jk} U_i^\dag - \delta_{ik} U_j^\dag \right)
\end{aligned}
\qquad \qquad \qquad
\begin{aligned}
\commute{M_0}{W_i^\dag} &= \frac{1}{2} \, W_i^\dag
\\
\commute{M_+}{W_i^\dag} &= 0
\\
\commute{M_-}{W_i^\dag} &= W_i
\\
\commute{L_{ij}}{W_k^\dag} &= i \left( \delta_{jk} W_i^\dag - \delta_{ik} W_j^\dag \right)
\end{aligned}
\end{equation}
and  close into the generators of $SO(d-2)_L \times SU(1,1)_M$:
\begin{equation}
\begin{aligned}
\commute{U_i}{W_j}
&= 2 i \, \delta_{ij} \, M_-
\\
\commute{U_i^\dag}{W_j^\dag}
&= 2 i \, \delta_{ij} \, M_+
\end{aligned}
\qquad \qquad
\begin{aligned}
\commute{U_i^\dag}{W_j}
&= \delta_{ij} \,\left( 2 i \, M_0 - \Delta \right)
+ 2 \, L_{ij}
\\
\commute{U_i}{W_j^\dag}
&= \delta_{ij} \,\left( 2 i \, M_0 + \Delta \right)
- 2 \, L_{ij}
\end{aligned}
\end{equation}
For the minimal unitary realization given above the quadratic Casimir of $\mathfrak{so}(d,2)$  turns out to be
\begin{equation}
\mathcal{C}_2 \left[ \mathfrak{so}(d,2) \right] = - \frac{1}{2} \left( d^2 -  4 \right) \,.
\end{equation}

The Lie algebra of $SO(d,2)$ admits a different 5-grading determined by the compact generator $M_0$  in addition to the 5-grading determined by the non-compact generator $\Delta$. The resulting $5\times 5$ grading is given in Table 1. We should note that the generators of $SO(d,2)$ given above are not all Hermitian. Since we are interested in unitary realizations one must go to a basis in which  all the generators at Hermitian ( anti-Hermitian) with pure imaginary (real) structure constants and determine the real form corresponding to the unitary realization , which  in our case turns out to be $SO(d,2)$. 

\begin{table}
	\begin{equation} \nonumber
	\begin{array}{ccccc}
	\phantom{R}~~~ & \phantom{U}~~~ &  K_- ~~~ & \phantom{V}~~~ & \phantom{R}~~~  \\[4pt]
	\phantom{R} & U_i & \vline \phantom{K} & U_i^\dagger  &\phantom{R} \\[4pt]
	M_-  &  --- & \left(  \Delta \oplus L_{ij} \oplus M_0  \right) &
	--- & M_+\\[4pt]
	\phantom{R} & W_i  & \vline \phantom{K} & W_i^\dagger & \phantom{R}  \\[4pt]  
	\phantom{R} & \phantom{W_i} & ~ K_+ ~~ & \phantom{W_i^\dagger} & \phantom{M_+} \\ [4pt]
	\end{array}
	\end{equation}
	\caption[]{
		Above we give the $5 \times 5$ grading of the  Lie
		algebra of $\mathfrak{so}(d,2)$. The vertical 5-grading is determined by
		$\Delta$ and the horizontal 5-grading is determined by
		$M_0$.}
	
\end{table}
The little group of massless particles in $d$ dimensional Minkowskian spacetime is the Euclidean group $ E_{(d-2)}$ in $(d-2)$ dimensions acting on the transverse coordinates whose Lie algebra is the semi-direct sum 
\[\mathfrak{e}_{(d-2)}=\mathfrak{so}(d-2) \, \circledS \, t_{(d-2)}\]
 where $t_{(d-2)}$ denotes the generators of translations.  There are two different such embeddings of the Euclidean group $E_{(d-2)}$ , namely  one acting on the transverse coordinates $X_i$ and the other  acting on the transverse momenta $P_i$ inside the quasiconformal realization of  $SO(d,2)$ which we shall denote as $E^X_{(d-2)}$ and $E^P_{(d-2)}$, respectively. Their generators are 
 \bea 
  E^X_{(d-2)} \Longrightarrow [L_{ij} \oplus  \, T^X_i  ] \\
  E^P_{(d-2)} \Longrightarrow  [ L_{ij} \oplus \, T^P_i  ]
  \eea
  where \[ T^X_i = \frac{i}{\sqrt{2}} (U_i^\dagger -U_i)= \, x P_i \]  and \[T^P_i = \frac{1}{\sqrt{2}} (U_i^\dagger +U_i)=  \, x X_i \] 
 Both of these Euclidean subgroups  have extensions to  conformal groups $SO(d-1,1)$  in $(d-2)$ Euclidean dimensions  as subgroups of $SO(d,2)$ , which we denote as $SO^X(d-1,1)$ and $SO^P(d-1,1)$. The Lie algebras of these Euclidean conformal subgroups in $(d-2)$ dimensions have the standard three graded decompositions:
 \bea
 \mathfrak{so}^X(d-1,1) = K_i^X \oplus ( L_{ij} + D^X ) \oplus T_i^X \\
 \mathfrak{so}^P(d-1,1) = K_i^P \oplus ( L_{ij} + D^P ) \oplus T_i^P  
 \eea
 where $D^X$ and $D^P$ are the the generators of respective scale transormations that determine the 3-grading given by 
 \bea  D^X &= \frac{1}{2} [ \Delta -i (M_+ -M_-)]= \frac{1}{2} ( xp+px) +\frac{1}{2}(X_i P_i +P_i X_i )  \\  D^P & = \frac{1}{2} [ \Delta +i (M_+ -M_-)]=\frac{1}{2} ( xp+px) - \frac{1}{2}(X_i P_i +P_i X_i )   \eea
  The special conformal generators $ K^X_i $ and $K^P_i$ acting on transverse coordinates $X_i$  and momenta $P_i$ are given by
  \bea  K^X_i = &\frac{1}{\sqrt{2}} (W_i +W_i^\dagger)= p \,  X_i - \frac{i}{x} \left( \frac{(d-3)}{2} X_i + i P_i X_j X_j - i P_jX_j X_i \right)
  \\ K^P_i= &\frac{i}{\sqrt{2}}(W_i^\dagger -W_i)=  p \, P_i - \frac{i}{x} \left( \frac{(d-3)}{2} P_i - i X_i P_j P_j + i X_jP_j P_i \right)\eea 
 
 They satisfy the commutation relations 
 \bea
 [T_i^X, K_j^X] &=&  -2i \delta_{ij} D^X + 2i L_{ij} \\  
 {[}T_i^P, K_j^P {]} &=&  2i \delta_{ij} D^P - 2i L_{ij}\\
 {[}T_i^X , K_j^P{]} &=& i L_+ \\
{[}T_i^P , K_j^X{]} &=& i L_- \\
 \eea
 where $L_+= P_i P_i $ and $L_-= X_i X_i $. 
We have a 5-grading of the Lie algebra $\mathfrak{so}(d,2)$ with respect to the generator $L_0=\frac{1}{2}(D^X-D^P)= X_i P_i +P_i X_i $ which together with $\Delta$ defines a $5\times 5$ grading  as indicated in Table 2.

\begin{table}
	\begin{equation} \nonumber
	\begin{array}{ccccc}
	\phantom{R}~~~ & \phantom{U}~~~ &  K_- ~~~ & \phantom{V}~~~ & \phantom{R}~~~  \\[4pt]
	\phantom{R} & T^P_i & \vline \phantom{K} & T_i^X  &\phantom{R} \\[4pt]
	L_-  &  --- & \left(  \Delta \oplus L_{ij} \oplus L_0  \right) &
	--- & L_+\\[4pt]
	\phantom{R} & K^X_i  & \vline \phantom{K} & K_i^P & \phantom{R}  \\[4pt]  
	\phantom{R} & \phantom{W_i} & ~ K_+ ~~ & \phantom{W_i^\dagger} & \phantom{M_+} \\ [4pt]
	\end{array}
	\end{equation}
	\caption[]{
		 Above we give the $5 \times 5$ grading of the  Lie
		algebra of $\mathfrak{so}(d,2)$ in an Hermitian basis. The vertical 5-grading is determined by
		$\Delta$ and the horizontal 5-grading is determined by
		$L_0$.}
\end{table}

Therefore the quasiconformal realization of $SO(d,2)$ can be interpreted as the minimal  Lie algebra containing the Euclidean conformal Lie algebra acting on transverse coordinates and the dual Euclidean conformal Lie algebra acting on the corresponding transverse momenta. The common subgroup of these two Euclidean conformal groups is $SO(d-2)$.

\section{ 3-grading of $\mathfrak{so}(d,2)$ as a conformal algebra in $d$-dimensions}
\label{sec:SO(d,2)nc3G}

The conformal group $SO(d,2)$ in $d$-dimensional Minkowskian space-time are generated by translations ($\mathcal{P}_\mu$ ), special conformal generators ($\mathcal{K}_\mu$ ) and Lorentz ($\mathcal{M}_{\mu\nu}$ ) and scale  tranformations ($\mathcal{D}$). Its Lie algebra $\mathfrak{so}(d,2)$ has a 3-grading determined by the generator $\mathcal{D}$ of scale transformations ( dilatations): 
\begin{equation}
	\mathfrak{so}(d,2)
	= \mathcal{K}_\mu \oplus
	( \mathcal{M}_{\mu\nu} + \mathcal{D} )  \oplus
	\mathcal{P}_\mu 
\end{equation}
In terms of the generators given in the previous section the  dilatation generator is given by
\begin{equation}
	\mathcal{D}
	= \frac{1}{2} \left[ \Delta - i \left( M_+ - M_- \right) \right] \,.
	\label{Dilatation}
\end{equation}
and the  Lorentz  generators $\mathcal{M}_{\mu\nu}$ ($\mu,\nu = 0,\dots,d-1$) are 
\begin{equation}
	\begin{split}
		\mathcal{M}_{0i}
		= \frac{1}{2\sqrt{2}} \left( U_i + U_i^\dag \right)
		+ \frac{i}{2\sqrt{2}} \left( W_i - W_i^\dag \right), 
		& \qquad
		\mathcal{M}_{ij}
		= L_{ij}
		\\
		\mathcal{M}_{i,d-1}
		= \frac{1}{2\sqrt{2}} \left( U_i + U_i^\dag \right)
		- \frac{i}{2\sqrt{2}} \left( W_i - W_i^\dag \right),
		& \qquad
		\mathcal{M}_{0,d-1}
		= \frac{1}{2} \left[ \Delta + i \left( M_+ - M_- \right) \right]
	\end{split}
	\label{lorentz}
\end{equation}
where $i,j,...=1,2,...,(d-2)$. 

The translation  $\mathcal{P}_\mu$ and special conformal generators $\mathcal{K}_\mu$ ($\mu = 0,\dots,d-1$) are given by
\begin{equation}
	\begin{split}
		\mathcal{P}_0
		&= K_+ + M_0 + \frac{1}{2} \left( M_+ + M_- \right)
		\\
		\mathcal{P}_i
		&= \frac{1}{\sqrt{2}} \left( W_i + W_i^\dag \right) \qquad \qquad (i = 1,\dots,d-2)
		\\
		\mathcal{P}_{d-1}
		&= K_+ - M_0 - \frac{1}{2} \left( M_+ + M_- \right)
	\end{split}
\end{equation}

\begin{equation}
	\begin{split}
		\mathcal{K}_0
		&= K_- + M_0 - \frac{1}{2} \left( M_+ + M_- \right)
		\\
		\mathcal{K}_i
		&= - \frac{i}{\sqrt{2}} \left( U_i - U_i^\dag \right) \qquad \qquad (i = 1,\dots,d-2)
		\\
		\mathcal{K}_{d-1}
		&= - K_- + M_0 - \frac{1}{2} \left( M_+ + M_- \right) \,.
	\end{split}
\end{equation}

They satisfy the commutation relations:
\begin{equation}
	\begin{split}
		\commute{\mathcal{M}_{\mu\nu}}{\mathcal{M}_{\rho\tau}}
		&= i \left(
		\eta_{\nu\rho} \mathcal{M}_{\mu\tau}
		- \eta_{\mu\rho} \mathcal{M}_{\nu\tau}
		- \eta_{\nu\tau} \mathcal{M}_{\mu\rho}
		+ \eta_{\mu\tau} \mathcal{M}_{\nu\rho}
		\right)
		\\
		\commute{\mathcal{P}_\mu}{\mathcal{M}_{\nu\rho}}
		&= i \left( \eta_{\mu\nu} \, \mathcal{P}_\rho
		- \eta_{\mu\rho} \, \mathcal{P}_\nu
		\right)
		\\
		\commute{\mathcal{K}_\mu}{\mathcal{M}_{\nu\rho}}
		&= i \left( \eta_{\mu\nu} \, \mathcal{K}_\rho
		- \eta_{\mu\rho} \, \mathcal{K}_\nu
		\right)
		\\
		\commute{\mathcal{D}}{\mathcal{M}_{\mu\nu}}
		&= \commute{\mathcal{P}_\mu}{\mathcal{P}_\nu}
		= \commute{\mathcal{K}_\mu}{\mathcal{K}_\nu}
		= 0
		\\
		\commute{\mathcal{D}}{\mathcal{P}_\mu}
		&= + i \, \mathcal{P}_\mu
		\qquad \qquad
		\commute{\mathcal{D}}{\mathcal{K}_\mu}
		= - i \, \mathcal{K}_\mu
		\\
		\commute{\mathcal{P}_\mu}{\mathcal{K}_\nu}
		&= 2 i \left( \eta_{\mu\nu} \, \mathcal{D} + \mathcal{M}_{\mu\nu} \right)
	\end{split}
\end{equation}
where $\eta_{\mu\nu} = \mathrm{diag} (-,+,\dots,+)$ is the Minkoswki metric in $d$ dimensions. 
The Poincar\'e mass operator in $d$ dimensions vanishes identically
\begin{equation}
	\mathscr{M}^2 = \eta_{\mu\nu} \mathcal{P}^\mu\mathcal{P}^\nu
	= 0
\end{equation}
 Similarly one finds that the square of special conformal generator $\mathcal{K}_\mu$ 
 vanishes identically as well
\begin{equation}
	\eta^{\mu\nu} \mathcal{K}_\mu \mathcal{K}_\nu = 0 \,.
\end{equation}

\section{3-grading of $\mathfrak{so}(d,2)$ with respect to its maximal compact subalgebra $\mathfrak{so}(d) \oplus \mathfrak{so}(2) $}
\label{sec:SO(d,2)c3G}

The generators $C_{MN} \oplus H$ ($M,N = 1,\dots,d$) of the maximal compact subalgebra $\mathfrak{so}(d) \oplus \mathfrak{so}(2)$  of
 $\mathfrak{so}(d,2)$ can be expressed in terms of the Lorentz covariant  generators $\mathcal{M}_{\mu\nu}, \mathcal{D} , \mathcal{K}_\mu$ and $\mathcal{P}_\mu $ as follows

\begin{equation}
	H = \frac{1}{2} ( \mathcal{P}_0 + \mathcal{K}_0 )= \frac{1}{2} \left( K_+ + K_- \right) + M_0 \,.
	\label{SO(2)generator}
\end{equation}

\begin{equation}
	\begin{aligned}
		C_{ij}
		&= L_{ij}
		\\
		C_{d-1,d}
		&=\frac{1}{2} \left( \mathcal{P}_{d-1} - \mathcal{K}_{d-1} \right)
	\end{aligned}
	\qquad \qquad
	\begin{aligned}
		C_{i,d-1}
		&= \mathcal{M}_{i,d-1} \\
		C_{i,d}
		&=\frac{1}{2} \left( \mathcal{P}_i - \mathcal{K}_i \right)
	\end{aligned}
	\label{SO(d)generators}
\end{equation}
They satisfy the commutation relations
\begin{equation}
	\commute{C_{MN}}{C_{PQ}}
	= i \left(
	\delta_{NP} C_{MQ} - \delta_{MP} C_{NQ}
	- \delta_{NQ} C_{MP} + \delta_{MQ} C_{NP}
	\right) \,.
\end{equation}

The Lie algebra $\mathfrak{so}(d,2)$ admits a "compact"  3-grading determined by the $SO(2)$ generator $H$ ( conformal hamiltonian):
\begin{equation}
\mathfrak{so}(d,2)
= C_M^- \oplus
( C_{MN} + H ) \oplus
C_M^+ \,.
\end{equation}
where the grade $+1$ generators $C_M^{+}$ are given by 
\begin{equation}
	\begin{split}
		C_i^+
		&= \mathcal{M}_{0i}-\frac{i}{2} (\mathcal{K}_i+\mathcal{P}_i) = \frac{1}{\sqrt{2}} \left( U_i^\dag - i \, W_i^\dag \right)
		\qquad \qquad (i = 1,\dots,d-2)
		\\
		C_{d-1}^+
		&= \mathcal{M}_{0,(d-1)} -\frac{i}{2} ( \mathcal{P}_{(d-1)} + \mathcal{K}_{(d-1)})= \frac{1}{2} \left[ \Delta - i \left( K_+ - K_- \right) \right] + i \, M_+
		\\
	C_d^+
		&= \frac{1}{2} (\mathcal{P}_0 -\mathcal{K}_0) + i \mathcal{D} = \frac{i}{2} \left[ \Delta - i \left( K_+ - K_- \right) \right] + M_+ \,.
	\end{split}
	\label{SO(d,2)3gradingC+}
\end{equation}
The grade $-1$ generators $C_M^-$  ($M,N,\dots = 1,\dots,d$) are Hermitian conjugates of the grade $+1$ generators $C_M^+$ : 
\begin{equation}
\begin{split}
C_i^-
&= \mathcal{M}_{0i}+\frac{i}{2} (\mathcal{K}_i+\mathcal{P}_i) = \frac{1}{\sqrt{2}} \left( U_i + i \, W_i \right)
\qquad \qquad (i = 1,\dots,d-2)
\\
C_{d-1}^-
&= \mathcal{M}_{0,(d-1)} +\frac{i}{2} ( \mathcal{P}_{(d-1)} + \mathcal{K}_{(d-1)})= \frac{1}{2} \left[ \Delta + i \left( K_+ - K_- \right) \right] - i \, M_-
\\
C_d^-
&= \frac{1}{2} (\mathcal{P}_0 -\mathcal{K}_0) - i \mathcal{D} = - \frac{i}{2} \left[ \Delta + i \left( K_+ - K_- \right) \right] + M_- \,.
\end{split}
\label{SO(d,2)3gradingC-}
\end{equation}
They satisfy 
\begin{equation}
	\begin{split}
		\commute{C_M^+}{C_{NP}}
		&= i \left(
		\delta_{MN} \, C_P^+ - \delta_{MP} \, C_N^+
		\right)
		\\
		\commute{C_M^-}{C_{NP}}
		&= i \left(
		\delta_{MN} \, C^-_P - \delta_{MP} \, C^-_N
		\right)
		\\
		\commute{H}{C_{MN}}
		&= \commute{C_M^+}{C_N^+}
		= \commute{C^-_M}{C^-_N}
		= 0
		\\
		\commute{H}{C_M^+}
		&= + C_M^+
		\qquad \qquad
		\commute{H}{C_M^-}
		= - C_M^-
		\\
		\commute{C_M^+}{C_N^-}
		&= 2  \left( - \delta_{MN} \, H + i \, C_{MN} \right)
	\end{split}
	\label{SO(d,2)Compact}
\end{equation}

\section{Hilbert Space of the Minimal Unitary Representation}
\label{sec:SU(1,1)_K}

The conformal Hamiltonian ( or AdS energy operator) $H$, given in equation (\ref{SO(2)generator}), can be written as the sum of  Hamiltonians $H_i$   of $(d-2)$ bosonic oscillators ($a_i$) and the Hamiltonian $H_\odot $ of  a singular oscillator:
\begin{equation}
\begin{split}
H &= \frac{1}{2} \left( K_+ + K_- \right) + M_0
\\
&= \frac{1}{4} \left( x^2 + p^2 \right)
+ \frac{1}{2 \, x^2} \mathcal{G}
+ \frac{1}{4}\sum_i  \, \left( a_i^\dag a_i + a_i a_i^\dag \right)
\\
&= H_\odot + \sum_i H_i
\end{split}
\end{equation}
where
\begin{equation}
H_\odot = \frac{1}{2} \left( K_+ + K_- \right)
= \frac{1}{4} \left( x^2 + p^2 \right) + \frac{1}{2 \, x^2} \mathcal{G}
\qquad ,\qquad
H_i
= \frac{1}{2} \, a_i^\dag a_i + \frac{1}{4}  \quad , \mathrm{ (no \, sum )} \,
\end{equation}
 $H_\odot$ is  the Hamiltonian of a singular harmonic oscillator. The role of the coupling constant for the singular potential 
\begin{equation}
V \left( x \right) = \frac{\mathcal{G}}{x^2}
\end{equation}
is played by the operator $\mathcal{G}$ given in equation (\ref{IsotonicCouplingConstant}).  The   Hamiltonian has the same form as  in  conformal quantum mechanics of  \cite{deAlfaro:1976je} as well as in  the Calogero models  \cite{Calogero:1969af,Calogero:1970nt}. It generates the compact $U(1)$ subgroup of the one-dimensional conformal group $SO(2,1)=SU(1,1)_K$ acting on the singlet coordinate $x$. The other generators of $SU(1,1)_K$ in the compact three grading with respect to $H_\odot$ are given by the following linear combinations   

\begin{equation}
\begin{split}
B^+_\odot
& 
= - \frac{i}{2} \left[ \Delta - i \left( K_+ - K_- \right) \right]
\\
B^-_\odot
&
= \frac{i}{2} \left[ \Delta + i \left( K_+ - K_- \right) \right]
\end{split}
\end{equation}
They satisfy the commutation relations:
\begin{equation}
\commute{B^-_\odot}{B_\odot^+} = 2 \, H_\odot
\qquad \qquad
\commute{H_\odot}{B_\odot^+} =  \, B_\odot^+
\qquad \qquad
\commute{H_\odot}{B^-_\odot} = - \, B^-_\odot
\end{equation}

As a basis for the Hilbert space of the minimal unitary representation of $SO(d,2)$ we shall consider states which are "twisted tensor products" of the states of the Fock space $\mathcal{F}$  of $(d-2)$ bosonic oscillators $a_i$ with the states of the singular oscillator that furnish a unitary representation of $SU(1,1)$\footnote{We use the term twisted tensor product since the eigenvalues of the Hamiltonian of the singular oscillator depend on the states of the Fock space $\mathcal{F}$.} . 
The Fock space of the $a$-type oscillators is spanned by 
the states of the form
\begin{equation}
\ket{n_1,n_2,\dots,n_{d-2}}
= \prod_i \frac{1}{\sqrt{n_{i} !}} \, ( \, a_i^\dag \, )^{n_{i}} \ket{0}
\end{equation}
where $n_{i}$ are non-negative integers and 
 $\ket{0}$  is the corresponding Fock vacuum.

In the coordinate representation the state(s) with the lowest eigenvalue of the conformal Hamiltonian $H_\odot$  are wave-functions of the form 
\be \psi_0^{\alpha_{g_0}} \left( x \right)
= C_0 \, x^{\alpha_{g_0}} e^{-x^2/2} \ee
where $C_0$ is a normalization constant  ,$g_0$ is the lowest eigenvalue of $\mathcal{G}$ and
\begin{equation}
\alpha_{g_0}
= \frac{1}{2} \pm \sqrt{2 g_0 + \frac{1}{4}} \,.
\end{equation}
The Fock vacuum $|0\rangle$ is the eigenstate of $\mathcal{G}$ with the lowest eigenvalue $g_0$. For the minrep of $SO(d,2)$ given above, the lowest possible value of $g_0$ is
\begin{equation}
g_0 = \frac{1}{8} \left( d - 3 \right) \left( d - 5 \right)
\end{equation} Denoting the state corresponding to the wave function $\psi_0^{\alpha_{g_0}} \left( x \right) $ with the Fock vacuum $|0\rangle$ 
 as $|\psi_0^{\alpha_{g_0}},0\rangle$ we have 
\begin{equation}
B^-_\odot \, | \psi_0^{\alpha_{g_0}},0\rangle  
= 0 \,.
\end{equation}
The Hermiticity of $H_\odot$ implies
\begin{equation}
g_0 \geq - \frac{1}{8}
\end{equation}
and the normalizability of $| \psi_0^{\alpha_{g_0}},0\rangle $ requires
\begin{equation}
\alpha_{g_0} > - \frac{1}{2} \,.
\end{equation}
 The lowest eigenvalue $g_0$ leads to two possible values  for $\alpha_{g_0}$ namely $\left( 5 - d \right) / 2$ and $\left( d - 3 \right) / 2$. However, the normalizability of the states in representation space requires one to choose \cite{Fernando:2015tiu}:
 \begin{equation}
 \alpha_{g_0} = \frac{(d-3)}{2} \,.
 \end{equation}

The corresponding tensor product state $
|\psi_0^{(d-3)/2}  \,,  0 \rangle
$
is an eigenstate of $H_\odot$ with the lowest eigenvalue $ E_\odot^{\alpha_{g_0}}=  \frac{1}{4} \left( d - 2 \right)$ :
\begin{equation}
H_\odot |\psi_0^{(d-3)/2}  \,,  0 \rangle = \frac{1}{4} \left( d - 2 \right)|\psi_0^{(d-3)/2} \,,  0 \rangle \,.
\end{equation}

The state $|\psi_0^{(d-3)/2} \left( x \right) \,,  0 \rangle $ is the unique state annihilated by all the grade $-1$ generators $ C^-_M$ and transforms as a singlet of $SO(d)$ subgroup. This shows that the minrep is a unitary lowest weight ( positive energy) representation. Furthermore it is annihilated by all the translation generators $U_i$ of the little group $SO(d-2)\circledS T_{(d-2)}$ of massless particles in $d$ dimensions. Therefore the minrep of $SO(d,2)$ describes a massless conformal  scalar field in $d$ dimensional Minkowskian spacetime. We refer to \cite{Fernando:2015tiu} for the K-type decomposition of the minrep of $SO(d,2)$  with respect to its maximal compact subgroup.

\section{Deformations of the minimal unitary representation of $SO(d,2)$ and massless conformal fields}
\label{sec:deformedSO(d,2)}
The minrep of $SO(d,2)$ describes a conformally massless scalar field in $d$ dimensions. In this section we shall review the "deformations" of the minrep and establish their one-to-one correspondence  with conformally massless fields in $d$ dimensional Minkoswkian spacetimes\cite{Fernando:2015tiu}. 
 By a deformation we mean adding spin terms $S_{ij}$ of the orbital generators $L_{ij}$ of the little group $SO(d-2)$ of massless particles such that all the Jacobi identities are satisfied.\footnote{ For $d=4$ the little group is $U(1)$ and deformations are parametrized by helicity which can take on continuous values \cite{Fernando:2009fq}. This explains the origin of the term "deformation" which has been used in all dimensions later.}:
\begin{equation}
L_{ij} \longrightarrow J_{ij}= L_{ij} + S_{ij}
\label{SO(d)J}
\end{equation}
 This requirement leads to following expression for  the generator  $K_+$ of the deformed minrep:
\begin{equation}
K_+ = \frac{1}{2} p^2
+ \frac{1}{x^2} \left(
\frac{1}{2} \, \mathcal{J}^2 - \frac{1}{4} \, \mathcal{L}^2
- \frac{(d-6)}{2(d-2)} \, \mathcal{S}^2
+ \frac{1}{8} \left( d - 3 \right) \left( d - 5 \right)
\right) \,.
\label{K+}
\end{equation}
where $\mathcal{J}^2= J_{ij} J_{ij} $ and $\mathcal{S}^2 = S_{ij} S_{ij} $. 
Under the deformation  the generators $M_{\pm,0}$ and $\Delta$ in grade 0 subspace, $U_i$ and $U_i^\dag$ in grade $-1$ subspace, and $K_-$ in grade $-2$ subspace of $\mathfrak{so}(d,2)$ remain unchanged. The grade $+1$ generators $W_i$ and $W_i^\dag$ get modified as follows:  :
\begin{equation}
\begin{split}
W_i
&= p \, a_i
- \frac{i}{x}
\left[
\frac{1}{2} \left( d - 3 \right) a_i + i \left( L_{ij} + 2  \, S_{ij} \right) a_j
\right]
\\
W_i^\dag
&= p \, a_i^\dag
- \frac{i}{x}
\left[
\frac{1}{2} \left( d - 3 \right) a_i^\dag + i \left( L_{ij} + 2  \, S_{ij} \right) a_j^\dag
\right]
\end{split}
\label{Grade+1Bosonic}
\end{equation}
Furthermore the spin generators $S_{ij}$ are required to satisfy the constraint: 
\begin{equation}
\Delta_{ij} = S_{ik} S_{jk} + S_{jk} S_{ik} - \frac{2}{(d-2)} \, \mathcal{S}^2 \, \delta_{ij} = 0
\label{SO(d-2)_Sconstraint}
\end{equation}
which turn out to be  the same constraint satisfied by  little group generators  of massless representations of the Poincar\'{e} group  that extend to unitary  representations of the conformal group in $d$ dimensions \cite{Angelopoulos:1997ij,Laoues:1998ik}. This  correspondence between massless conformal  fields in $d$ dimensions and the minrep of $SO(d,2)$ and its deformations were established earlier for dimensions $d=4, 5$ and $6$  in  \cite{Fernando:2009fq,Fernando:2010dp,Fernando:2010ia,Fernando:2014pya}. More recently  this one-to-one correspondence was extended to all spacetime dimensions in \cite{Fernando:2015tiu}. We should note that  each deformation leads to a different unitary realization of the $SU(1,1)_K$ subgroup and  thus establishes a correspondence between massless conformal fields in any dimension and the Calogero models.

  For the deformed minrep one finds the following expressions for the Casimirs of various subalgebras 
\begin{equation}
\begin{split}
\mathcal{C}_2 \left[ \mathfrak{so}(d-2)_J \right]
&= J_{ij} J_{ij}
= \mathcal{J}^2
\\
\mathcal{C}_2 \left[ \mathfrak{su}(1,1)_M \right]
&= {M_0}^2 - \frac{1}{2} \left( M_+ M_- + M_- M_+ \right)
= \mathcal{M}^2
= \frac{1}{8} \, \mathcal{L}^2 + \frac{1}{16} \left( d - 2 \right) \left( d - 6 \right)
\\
\mathcal{C}_2 \left[ \mathfrak{su}(1,1)_K \right]
&= \Delta^2 - 2 \left( K_+ K_- + K_- K_+ \right)
= \mathcal{K}^2
\\
&= - \mathcal{J}^2 + \frac{1}{2} \, \mathcal{L}^2 + \frac{(d-6)}{(d-2)} \, \mathcal{S}^2 - \frac{1}{4} \left( d - 2 \right) \left( d - 6 \right)
\end{split}
\end{equation}
The coset $SO(d,2)/SO(d-2)\times SO(2,2)$ generators satisfy 
\begin{equation} 
\left[ U W \right]
= U_i W_i^\dag + W_i^\dag U_i - U_i^\dag W_i - W_i U_i^\dag
= 2 i \left( \mathcal{J}^2 - \mathcal{S}^2 \right)
+ i \left( d - 2 \right)^2
\end{equation}
leading to following expression for 
the quadratic Casimir of $\mathfrak{so}(d,2)$
\begin{equation}
\begin{split}
\mathcal{C}_2 \left[ \mathfrak{so}(d,2) \right]
&= \frac{(d+2)}{(d-2)} \, \mathcal{S}^2
- \frac{1}{2} \left( d^2 - 4 \right) \,.
\end{split}
\end{equation}
which agrees with the quadratic Casimir of the undeformed minrep of $\mathfrak{so}(d,2)$ when $\mathcal{S}^2 = 0$. The minrep and its deformations are unitary lowest weight ( positive energy) representations of $SO(d,2)$ that are uniquely determined by the lowest energy  irrep of $SO(d)$ subgroup. Therefore we shall label them  as
\[ | E_0, (n_1,\cdots n_r ) \rangle \]
where $E_0$ is the eigenvalue of $H$ and $(n_1, \cdots n_r)$ are the Dynkin labels of lowest energy irrep of $SO(d)$.

\subsection{Deformations in odd dimensions}
It is well-known that the  $AdS_4/CFT_3$ Lie algebra  $SO(3,2)$ is isomorphic to the symplectic Lie algebra $Sp(4,\mathbb{R})$ whose minimal unitary  representation that describes a conformal scalar  admits a single deformation which describes a conformal spinor in 3 dimensions. They are simply the scalar and spinor singletons of Dirac\cite{Dirac:1963ta} which describe the only conformally massless fields in $d=3$.  
Similarly, the minrep of $AdS_6/CFT_5$ group $SO(5,2)$ admits a unique deformation\cite{Fernando:2014pya}. The minrep of $SO(5,2)$ and its deformation  are the analogs of Dirac singletons  and decribe  conformally massless scalar and spinor  fields in $d=5$. This phenomenon  extends to all  conformal algebras $\mathfrak{so}(d,2)$ in odd dimensions as a consequence of  the fact that the  constraint \ref{SO(d-2)_Sconstraint} for  spin terms $S_{ij}$ of $SO(d-2)$  that determine the deformations of the minrep   has a unique non-trivial solution for odd $d$. It  is simply the $2^{\left( \frac{d-3}{2} \right)}$ dimensional  spinor  irrep of $SO(d-2)$ generated by
\begin{equation}
S_{ij}= \frac{1}{4} [ \gamma_i, \gamma_j] \,.
\end{equation}
 where $\gamma_i$ are the 
 $2^n\times 2^n$ ($n=(d-3)/2)$ Euclidean Dirac gamma matrices in $(d-2)$ dimensions. 
The minrep and its unique deformations describe the conformally massless scalar and spinor fields.  They exhaust the list of conformally masless fields for odd $d$\cite{Laoues:1998ik}. The lowest energy irrep of the minrep has the labels
\[ | \frac{(d-2)}{2}, (0,\cdots , 0)_D\rangle \]
while the unique spinorial deformation of the minrep has the lowest energy irrep
\[ | \frac{(d-1)}{2}, (0,\cdots , 0,1)_D\rangle \]
\subsection{Deformations in even dimensions}
The minrep of the conformal group in four dimensions admits a one parameter family of deformations labelled by helicity\cite{Fernando:2009fq}. For integer and half integer values of helicity the resulting unitary representations are isomorphic to the doubleton representations that describe massless conformal fields of arbitrary spin in four dimensions that were constructed and studied in\cite{Gunaydin:1984fk,Gunaydin:1998jc,Gunaydin:1998sw}. Similarly one finds that the minrep of $SO(6,2)$ admits a discrete   infinite family  of deformations\cite{Fernando:2010dp,Fernando:2010ia} that are isomorphic to doubleton representations that describe higher spin massless conformal fields in six dimensions  studied earlier in \cite{Gunaydin:1984wc,Gunaydin:1999ci,Fernando:2001ak}. 

In contrast to odd dimensions, the constraints on the spin terms that describe the deformations of the minrep of  $SO(d,2)$ for even $d$  admit an infinite set of solutions that describe conformally massless fields. They exhaust the list of conformally massless fields as a consequence of the fact that \ref{SO(d-2)_Sconstraint} is a necessary and sufficient condition for a massless representation of the Poincare group to be extendable  to a unitary representation of the conformal group\cite{Angelopoulos:1997ij,Laoues:1998ik}. These deformations are all unitary lowest weight representations of $SO(d,2)$ whose lowest energy irreps have  the $SO(2)\times SO(d)$ labels\cite{Fernando:2015tiu}
\[ | E_0=\frac{(d+s-2)}{2}; (0,\cdots,0, 0, s)_D\rangle \]  
or 
\[ | E_0=\frac{(d+s-2)}{2}; (0,\cdots,0,s, 0)_D\rangle \] 
where $s$ is a non-negative integer. 
The spinorial generators $S_{ij}$ ($i,j,\dots =1,\dots, (d-2) $ ) can always be realized in terms of fermionic oscillators\cite{Fernando:2015tiu}.
\section{Quasiconformal construction of  higher spin algebras} 
\subsection{Bosonic $AdS_{(d+1)}/CFT_d$  higher spin algebras}
\label{sec:bosonicHS}

To construct the $AdS{(d+1)}/CFT_d$ higher spin algebra , let us first express the $\mathfrak{so}(d,2)$ generators in the ``canonical basis'' $M_{AB}$ ($A,B=0,1,\dots,d+1$)  in terms of the generators of the compact 3-grading 
\begin{equation}
\begin{aligned}
M_{0M} &= \frac{1}{2} \left( C_M^+ + C_M^- \right)
\\
M_{MN} &=C_{MN}
\end{aligned}
\qquad \qquad \qquad
\begin{aligned}
M_{0,d+1} &= H
\\
M_{M,d+1} &= \frac{i}{2} \left( C_M^+ - C_M^- \right)
\end{aligned}
\end{equation}
They satisfy the  commutation relations:\begin{equation}
\commute{M_{AB}}{M_{CD}}
= i \left( \eta_{BC} M_{AD} - \eta_{AC} M_{BD} - \eta_{BD} M_{AC} + \eta_{AD} M_{BC} \right)
\end{equation}
where $\eta_{AB} = \mathrm{diag} \left( -,+,\dots,+,- \right)$ is the $SO(d,2)$ invariant metric.

In terms of the Lorentz covariant generators  $\mathcal{D}$, $\mathcal{M}_{\mu\nu}$, $\mathcal{P}_\mu$, $\mathcal{K}_\mu$ ($\mu,\nu=0,\dots,d-1$), the generators $M_{AB}$ take the form:
\begin{equation}
\begin{aligned}
M_{\mu\nu} &= \mathcal{M}_{\mu\nu}
\\
M_{\mu,d} &= \frac{1}{2} \left( \mathcal{P}_\mu - \mathcal{K}_\mu \right)
\end{aligned}
\qquad \qquad \qquad
\begin{aligned}
M_{\mu,d+1} &= \frac{1}{2} \left( \mathcal{P}_\mu + \mathcal{K}_\mu \right)
\\
M_{d,d+1} &= - \mathcal{D}
\end{aligned}
\end{equation}

The $AdS_{(d+1)}/CFT_d$ higher spin algebra is simply the quotient of  the universal enveloping algebra $ \mathscr{U}(d,2) $ of $SO(d,2)$ by its Joseph ideal $\mathscr{J}(d,2)$ \cite{Gunaydin:1989um,Vasiliev:1999ba,Eastwood:2002su,eastwood2005uniqueness,Govil:2013uta,Govil:2014uwa,Fernando:2014pya}.
We recall that the Joseph ideal of the universal enveloping algebra of a Lie algebra  is a two-sided ideal that annihilates its minimal unitary representation. In fact , in the mathematics literature minrep is defined by this property. 
We shall denote the corresponding $AdS_{(d+1)}/CFT_d$ higher spin algebra as $hs(d,2)$:
\begin{equation}
hs(d,2) = \frac{\mathscr{U}(d,2)}{\mathscr{J}(d,2)}
\end{equation}

In terms of  $M_{AB}$  the explicit expression for the generators $J_{ABCD}$  of the Joseph ideal of $SO(d,2)$ was given in \cite{eastwood2005uniqueness}:
\begin{equation}
\begin{split}
J_{ABCD}
&= M_{AB} M_{CD}
- M_{AB} \circledcirc M_{CD}
- \frac{1}{2} \commute{M_{AB}}{M_{CD}}
+ \frac{(d-2)}{4d(d+1)} \langle M_{AB} , M_{CD} \rangle \, \mathbf{1}
\\
&= \frac{1}{2} M_{AB} \cdot M_{CD}
- M_{AB} \circledcirc M_{CD}
+ \frac{(d-2)}{4d(d+1)} \langle M_{AB} , M_{CD} \rangle \,  \mathbf{1} \,.
\end{split}
\label{Joseph}
\end{equation}
where the symbol $\cdot$ denotes the symmetric product
\begin{equation}
M_{AB} \cdot M_{CD}
= M_{AB} M_{CD} + M_{CD} M_{AB} \,,
\end{equation}
and the symbol $\circledcirc$ denotes the Cartan product of two generators of $\mathfrak{so}(d,2)$ \cite{eastwood2005cartan}:
\begin{equation}
\begin{split}
M_{AB} \circledcirc M_{CD}
&= \frac{1}{3} M_{AB} M_{CD}
+ \frac{1}{3} M_{DC} M_{BA}
\\
& \quad
+ \frac{1}{6} M_{AC} M_{BD}
- \frac{1}{6} M_{AD} M_{BC}
+ \frac{1}{6} M_{DB} M_{CA}
- \frac{1}{6} M_{CB} M_{DA}
\\
& \quad
- \frac{1}{2d}
\left(
\eta_{BD} M_{AE} M_C^{~E} - \eta_{AD} M_{BE} M_C^{~E}
+ \eta_{AC} M_{BE} M_D^{~E} - \eta_{BC} M_{AE} M_D^{~E}
\right)
\\
& \quad
- \frac{1}{2d}
\left(
\eta_{BD} M_{CE} M_A^{~E} - \eta_{AD} M_{CE} M_B^{~E}
+ \eta_{AC} M_{DE} M_B^{~E} - \eta_{BC} M_{DE} M_A^{~E}
\right)
\\
& \quad
+ \frac{1}{d(d+1)} M_{EF} M^{EF}
\left( \eta_{AC} \eta_{BD} - \eta_{BC} \eta_{AD} \right) \,,
\end{split}
\end{equation}
The bilinear product $\langle M_{AB} , M_{CD} \rangle$ is defined by the Killing form of $SO(d,2)$:
\begin{equation}
\langle M_{AB} , M_{CD} \rangle
= - \frac{2d}{(d^2-4)} \, M_{EF} M_{GH}
\left( \eta^{EG} \eta^{FH} - \eta^{EH} \eta^{FG} \right)
\left( \eta_{AC} \eta_{BD} - \eta_{AD} \eta_{BC} \right)
\end{equation}

Under the  adjoint action of $\mathfrak{so}(d,2)$  the decomposition of the enveloping algebra $\mathscr{U}(d,2)$  is  given by the symmetrized products of the adjoint representation  with  Young tableau
\begin{equation}
M_{AB} \sim \parbox{10pt}{\YoungAA} \,\,\,.
\end{equation}
The symmetrized tensor product of two copies of the  adjoint representation  decomposes as:
\begin{equation}
\begin{split}
\left( \,\,\, \parbox{10pt}{\YoungAA} \,\, \otimes \,\, \parbox{10pt}{\YoungAA} \,\,\, \right)_S
&= \,\,\, \parbox{20pt}{\YoungBB}
\,\,\, \oplus \,\,\,
\parbox{10pt}{\YoungAAAA}
\,\,\, \oplus \,\,\,
\parbox{20pt}{\YoungB}
\,\,\, \oplus \,\,\,
\bullet
\end{split}
\label{adjointsymm}
\end{equation}
where $\bullet$ represents the quadratic Casimir of $SO(d,2)$. As was shown by Vasiliev \cite{Vasiliev:1999ba}, the higher spin gauge fields in $AdS_{(d+1)}$ transform in the  traceless two-row Young tableaux representations of $SO(d,2)$:
\begin{equation} 
\underbrace{\begin{picture}(100,30)(0,0)
	\put(0,5){\line(0,1){20}}
	\put(00,5){\line(1,0){100}}
	\put(00,15){\line(1,0){100}}
	\put(00,25){\line(1,0){100}}
	\put(10,5){\line(0,1){20}}
	\put(20,5){\line(0,1){20}}
	\put(30,5){\line(0,1){20}}
	\put(40,5){\line(0,1){20}}
	\put(70,5){\line(0,1){20}}
	\put(80,5){\line(0,1){20}}
	\put(90,5){\line(0,1){20}}
	\put(100,5){\line(0,1){20}}
	\put(55,10){\makebox(0,0){$\cdots$}}
	\put(55,20){\makebox(0,0){$\cdots$}}
	\end{picture}}_{\mbox{$n$ boxes}}
\end{equation}
Therefore one has to mod out all the representations on the right hand side of equation (\ref{adjointsymm}) except for the window diagram. Quotienting the enveloping algebra by the Joseph ideal does precisely this and the generators of the resulting infinite higher spin algebra transform  under $\mathfrak{so}(d,2)$ as follows:
\begin{equation}
\parbox{10pt}{\YoungAA} \quad \oplus \quad  \parbox{20pt}{\YoungBB} \quad \oplus \quad ...\quad \oplus\quad 
\begin{picture}(100,20)(0,12)
\put(0,5){\line(0,1){20}}
\put(00,5){\line(1,0){100}}
\put(00,15){\line(1,0){100}}
\put(00,25){\line(1,0){100}}
\put(10,5){\line(0,1){20}}
\put(20,5){\line(0,1){20}}
\put(30,5){\line(0,1){20}}
\put(40,5){\line(0,1){20}}
\put(70,5){\line(0,1){20}}
\put(80,5){\line(0,1){20}}
\put(90,5){\line(0,1){20}}
\put(100,5){\line(0,1){20}}
\put(55,10){\makebox(0,0){$\cdots$}}
\put(55,20){\makebox(0,0){$\cdots$}}
\end{picture} \quad \oplus \quad \dots
\end{equation}

{\it For the minimal unitary realization of  $SO(d,2)$ obtained via  the quasiconformal method  the  generators $J_{ABCD}$ of the Joseph ideal vanish identically as  operators}.  For $d=3,4,6$ dimensions this was established in \cite{Govil:2013uta,Govil:2014uwa}.  These results were extended to $d=5$ in \cite{Fernando:2014pya} and to general $d>6$ in\cite{Fernando:2015tiu}.
Hence  the enveloping algebra of the minrep of $SO(d,2)$ that results from quantization of its geometric quasiconformal action yields directly  the bosonic higher spin $AdS_{(d+1)}$/Conf$_d$ algebra in all dimensions. 

\subsection{Deformations and supersymmetric extensions of $AdS_{(d+1)}/CFT_d$  higher spin algebras} 
The minimal unitary representation of $SO(d,2)$ and its  deformations as obtained via the quasiconformal approach are in one-to-one correspondence with  conformally massless fields in $d$-dimensional Minkowskian spacetimes. Therefore one defines the  {\it deformations} of the bosonic $AdS_{(d+1)}/CFT_d $ higher spin algebras as enveloping algebras of the deformed minimal unitary representations\cite{Govil:2013uta,Govil:2014uwa,Fernando:2014pya,Fernando:2015tiu}. These deformed higher spin algebras can be interpreted as quotients of the universal enveloping algebra $\mathscr{U}(d,2) $ of $SO(d,2)$ by a deformed Joseph ideal $\mathscr{J}_{Def}(d,2)$:
\begin{equation}
hs(d,2)_{Def} = \frac{\mathscr{U}(d,2)}{\mathscr{J}_{Def}(d,2)}
\end{equation}
The deformations of the Joseph ideal  are best studied by  decomposing the generators $J_{ABCD}$ of the Joseph ideal with respect to the Lorentz group  $SO(d-1,1)$ as was done in \cite{Govil:2013uta,Govil:2014uwa,Fernando:2014pya,Fernando:2015tiu}. 
Vanishing of the generators $J_{ABCD}$  as operators within quasiconformal realization of  the minrep are equivalent to the following Lorentz covariant  conditions:
\begin{equation}
\begin{split}
d \, \mathcal{D} \cdot \mathcal{D} + \mathcal{M}^{\mu\nu} \cdot M_{\mu\nu} + \frac{(d-2)}{2} \, \mathcal{P}^\mu \cdot \mathcal{K}_\mu
&= 0
\\
\mathcal{P}^\mu \cdot \left( \mathcal{M}_{\mu\nu} + \eta_{\mu\nu} \, \mathcal{D} \right)
&= 0
\\
\mathcal{K}^\mu \cdot \left( \mathcal{M}_{\mu\nu} - \eta_{\mu\nu} \, \mathcal{D} \right)
&= 0
\\
\eta^{\mu\nu} \, \mathcal{M}_{\mu\rho} \cdot \mathcal{M}_{\nu\sigma} - \mathcal{P}_{(\rho} \cdot \mathcal{K}_{\sigma)} + \left( d - 2 \right) \eta_{\rho\sigma}
&= 0
\\
\mathcal{M}_{\mu\nu} \cdot \mathcal{M}_{\rho\sigma} + \mathcal{M}_{\mu\sigma} \cdot \mathcal{M}_{\nu\rho} + \mathcal{M}_{\mu\rho} \cdot \mathcal{M}_{\sigma\nu}
&= 0
\\
\mathcal{D} \cdot \mathcal{M}_{\mu\nu} + \mathcal{P}_{[\mu} \cdot \mathcal{K}_{\nu]}
&= 0
\\
\mathcal{M}_{[\mu\nu} \cdot \mathcal{P}_{\rho]}
&= 0
\\
\mathcal{M}_{[\mu\nu} \cdot \mathcal{K}_{\rho]}
&= 0
\end{split}
\end{equation}
where symmetrizations (round brackets) and anti-symmetrizations (square brackets) are of weight one.

Under deformation some of the generators of $SO(d,2)$ get modified by spin dependent terms. More specifically  only the first three identities above remain unchanged and 
the momentum and special conformal generators remain lightlike   under deformations:
\begin{equation}
P^2 = \mathcal{P}^\mu \mathcal{P}_\mu = 0
\quad , \quad
K^2 = \mathcal{K}^\mu \mathcal{K}_\mu = 0
\end{equation}
The fourth identity above takes the form\cite{Fernando:2015tiu}
\begin{equation}
\eta^{\mu\nu} \mathcal{M}_{\mu\rho} \cdot \mathcal{M}_{\nu\sigma}
- \mathcal{P}_{(\rho} \cdot \mathcal{K}_{\sigma)}
- \left[ \frac{2}{(d-2)} \mathcal{S}^2 - \left( d - 2 \right) \right] \eta_{\rho\sigma}
= 0
\end{equation}
where $\mathcal{S}^2$ is the Casimir operator of the little group  $SO(d-2)_S$ generated by the spin terms $S_{ij}$. 


Physical interpretation of some of the deformed identities  depends on the space-time dimension.
In four dimensions one finds that the identities that depend on the  deformation parameter $\zeta$ , which is twice the helicity, take the form\cite{Govil:2013uta}
\bea
W^\mu=\half \epsilon^{\mu\nu\rho\sigma}M_{\nu\rho} \cdot P_{\sigma} \eq \zeta P^\mu \\ V^\mu =
\half \epsilon^{\mu\nu\rho\sigma}M_{\nu\rho} \cdot K_{\sigma} \eq -\zeta K^\mu
\eea
\bea
 \eta^{\mu\nu} M_{\mu\rho} \cdot M_{\nu\sigma} -  P_{(\rho} \cdot K_{\sigma)} + 2 \eta_{\rho\sigma} \eq \frac{\zeta^2}{2} \eta_{\rho\sigma} \\
M_{\mu\nu} \cdot M_{\rho\sigma} + M_{\mu\sigma}\cdot M_{\nu\rho} + M_{\mu\rho}\cdot M_{\sigma\nu}   \eq  \zeta\epsilon_{\mu\nu\rho\sigma} \Delta \\
\Delta \cdot M_{\mu\nu} + P_{[\mu} \cdot K_{\nu]} \eq  - \frac{\mathcal{\zeta}}{2} \epsilon_{\mu\nu \rho\sigma}M^{\rho\sigma}
\eea
 and $W^\mu$ and $V^\mu$ are the Pauli-Lubansky vector and its special conformal analog, respectively.
The eigenvalues of the Casimir operators of the deformed  minreps of $SO(4,2)$  depend only on $\zeta$\cite{Govil:2013uta}:
\bea
C_2 &=& M^{~A}_B M^{~B}_A = 6 - \frac{3\zeta^2}{2} \\
C_3 & =&  \epsilon^{ABCDEF}M_{AB}M_{CD}M_{EF}= 6 \zeta \lb \zeta^2-4 \rb = -8C_2 \sqrt{1-\frac{C_2}{6}} \\
C_4 &= &M^{~A}_B M^{~B}_C M^{~C}_D M^{~D}_A= \frac{3}{8} \lb \zeta^4 + 8\zeta^2 -48 \rb = \frac{C_2^2}{6} - 4 C_2
\eea



Deformed Joseph ideal generators of $6d$ conformal group $SO(6,2)$ were given in\cite{Govil:2014uwa}.
One finds that the analog of Pauli-Lubanski vector in $6d$ is a tensor of rank 3 and its conformal analogue defined as
\be
A_{\mu\nu\rho} =\frac{1}{3!}  \epsilon_{\mu\nu\rho\sigma\delta\tau} M^{[\sigma\delta} \cdot P^{\tau]} \qquad B_{\mu\nu\rho} =\frac{1}{3!}  \epsilon_{\mu\nu\rho\sigma\delta\tau} M^{[\sigma \delta} \cdot K^{\tau]}
\ee
They  vanish identically for the minrep 
\be
A_{\mu\nu\rho} = 0 \qquad B_{\mu\nu\rho} = 0
\ee
For the deformed minreps they do not vanish, but satisfy self-duality  and anti-self-duality conditions: 
\bea
A_{\mu\nu\rho} \eq \widetilde{A}_{\mu\nu\rho} \nn
B_{\mu\nu\rho} \eq -\widetilde{B}_{\mu\nu\rho} 
\label{so62def2}
\eea
where the dual rank three tensors are defined as follows:
\be
\widetilde{A}_{\mu\nu\rho} = \frac{1}{3!}  \epsilon_{\mu\nu\rho\sigma\delta\tau} A^{\sigma\delta\tau}, \qquad \widetilde{B}_{\mu\nu\rho} = \frac{1}{3!}  \epsilon_{\mu\nu\rho\sigma\delta\tau} B^{\sigma\delta\tau}
\ee

Similarly one finds  the following identity  for the generators of the deformed minreps:
\be
M_{\mu\nu} \cdot M_{\rho\sigma} + M_{\mu\sigma}\cdot M_{\nu\rho} + M_{\mu\rho}\cdot M_{\sigma\nu}   = \epsilon_{\mu\nu\rho\sigma}^{\quad\,\,\,\,\, \delta\tau} (P_{[\delta} \cdot K_{\tau]} + M_{\delta\tau} \cdot \Delta)
\label{so62def1}
\ee
while each side of this equation vanishes separately for the true minrep.

The quadratic Casimir operator of the deformed minrep of $SO^*(8)$is given by  \cite{Fernando:2010dp}:
\be
C_2 \left[ \mathfrak{so}^*(8) \right]_{\text{deformed}} = - 8 \lb 2 -  \mathcal{T}^2 \rb
\ee
where $\mathcal{T}^2$ is the quadratic Casimir operator of an $SU(2)$ subgroup of the little  group $SO(4)$ of massless particles whose eigenvalues label the deformations.

In general the representation \[ \parbox{10pt}{\YoungAAAA} \] that vanished  in the symmetric tensor product of the adjoint representation for the true minrep of $SO(d,2)$ no longer vanishes for the deformed minrep. On the other hand the representation $\parbox{20pt}{\YoungB}$ that occurs in the tensor product  still vanishes for the deformed minreps:
\begin{equation}
	\eta^{CD} M_{AC} \cdot M_{DB} = 0
\end{equation}
Hence  the gauge fields of deformed higher spin theories defined by  $hs(d,2)_{Def} $  will not  consist only of  fields corresponding to traceless two-row Young tableaux. Their Young tableaux will be  those  that occur in  the symmetrized  products of the adjoint tableau $\parbox{10pt}{\YoungAA}$ of $SO(d,2)$ subject to  the constraint:
\begin{equation*}
	\left( \,\, \parbox{10pt}{\YoungAA} \, \otimes \, \parbox{10pt}{\YoungAA} \,\, \right)_{Sym}
	= \,\, \parbox{20pt}{\YoungBB} \, \oplus \, \parbox{10pt}{\YoungAAAA}
\end{equation*}

\subsection{Supersymmetric deformations of higher spin algebras}

Maximal dimension for the existence of simple superconformal algebras over the field of real or complex numbers is six\cite{Nahm:1977tg}. 
In $d=3$ the conformal group $SO(3,2)$ admits supersymmetric extensions to supergroups $OSp(N|4,\mathbb{R})$ with even subgroups $SO(N)\times Sp(4,\mathbb{R})$. The minimal unitary supermultiplet of $OSp(N|4,\mathbb{R})$ for even $N$ consists of a scalar singleton transforming in a chiral spinor representation of $SO(N)$ and a spinor singleton transforming in the conjugate spinor representation of $SO(N)$. For odd $N$ the minimal unitary supermultiplet contains a scalar and a spinor singleton both transforming in the unique spinor representation of $SO(N)$.

 That the higher spin algebra of Fradkin and Vasiliev in $AdS_4$ as well as its supersymmetric extensions are given by the enveloping algebras of the  singletonic realization of $SO(3,2)$ and its supersymmetric extensions was first pointed out in \cite{Gunaydin:1989um}. 
In higher dimensions $AdS_{(d+1)}/CFT_d$ higher spin superalgebras are defined as enveloping algebras of the quasiconformal  realization of the minimal unitary representation of the correponding superconformal algebras\cite{Govil:2013uta,Govil:2014uwa,Fernando:2014pya}. The minimal unitary supermultiplet contains the minrep of $SO(d,2)$ and certain deformations of the minrep. As such it corresponds to a massless conformal supermultiplet of the underlying conformal superalgebra.

In $d=4$ the conformal algebra $SU(2,2)$  can be extended to an infinite family of 
 superalgebras $SU(2,2\,|\,N)$ with the even subalgebra $SU(2,2) \oplus U(N)$.
 The minimal unitary representation of $SU(2,2|N )$ and its deformations were constructed via the quasiconformal method  in \cite{Fernando:2009fq}. These representations were later reformulated in terms of helicity deformed twistors that transform non-linearly under the Lorentz group and the corresponding $AdS_5/CFT_4$ higher spin superalgebras were studied in \cite{Govil:2013uta}.  
 
The conformal algebra $SO(6,2)=SO^*(8)$ in $d=6$ also admits an infinite family of superconformal extensions  $OSp(8^*\,|\,2N)$ with even subalgebras $SO^*(8) \oplus USp(2N)$.  The   quasiconformal construction of the minimal unitary supermultiplets of $OSp(8^*\,|\,2N)$ and their deformations were given in \cite{Fernando:2010dp,Fernando:2010ia}.   These representaions were reformulated in terms of deformed twistors and the corresponding $AdS_7/CFT_6$ higher spin superalgebras were studied in \cite{Govil:2014uwa}.

In five dimensions there exists a unique superconformal algebra , namely the exceptional superalgebra with the even subalgebra $SO(5,2) \oplus SU(2)$. The minimal unitary supermultiplet of $F(4)$ was constructed via the quasiconformal method  and the unique higher spin $AdS_6/CFT_5$ superalgbera was studied in \cite{Fernando:2015tiu}.

{\bf Acknowledgements:}  I would like to thank Sudarshan Fernando, Karan Govil and Oleksandr Pavlyk for  most enjoyable and stimulating scientific collaborations on topics covered in this review.  I would also like to thank Evgeny Skvortsov, Massimo Taronna and  Misha Vasiliev for many helpful discussions on higher spin theories. Thanks are also due to Lars Brink and other organizers for the invitation to  the Conference  on Higher Spin Gauge Theories at the Institute of Advanced Studies in Nanyang Technological University in Singapore. The research reviewed here was supported in part by the US Department of Energy under DOE Grant No: DE-SC0010534 and US National Science Foundation under grants PHY-1213183 and PHY-08-55356.
\providecommand{\href}[2]{#2}\begingroup\raggedright\endgroup
\end{document}